# Revealing compositional evolution of PdAu electrocatalyst by atom probe tomography

Leonardo Shoji Aota[a], Se-Ho Kim[a,b,*], Chanwon Jung[a], Siyuan Zhang[a], Baptiste Gault[a,c,*]

[a] Max-Planck-Institut für Eisenforschung, Max-Planck-Straße 1, 40237 Düsseldorf, Germany
[b] Department of Materials Science and Engineering, Korea University, Seoul 02841, Republic of Korea
[c] Department of Materials, Imperial College London, SW7 2AZ London, UK
[*]Corresponding authors

## Abstract

Pd-based electro-catalysts are a key component to improve the methanol oxidation reaction (MOR) kinetics from alcohol fuel cells. However, the performance of such catalysts is degraded over time. To understand the microstructural/atomic scale chemical changes responsible for such effect, scanning (transmission) electron microscopies and atom probe tomography (APT) were performed after accelerated degradation tests (ADT). No morphological changes are observed after 1000 MOR cycles. Contrastingly, (1) Pd and B are leached from PdAu nanoparticles and (2) Au-rich regions are formed at the surface of the catalyst. These insights highlight the importance of understanding the chemical modification undergoing upon MOR to design superior catalysts.

Alcohol fuel cells (AFCs) are an attractive alternative to traditional proton electrolyte membrane fuel cells, owing to the ease of the fuel's storage and transportation, and, as such, are a potential candidate to replace current fossil fuel engines[1,2]. In AFCs, alcohol is used as a fuel to generate electricity rather than heat, which is generated by combustion[3,4]. As a result, a higher transformation efficiency is obtained while using an abundant and possibly sustainable fuel, generated through fermentation of biomass such as corn and other plants[3–7]. A remarkable effort has been devoted to the development of different AFCs based on methanol[8], ethanol[5,7,9], ethylene glycol[5,6,10], and glycerol[11]. Direct methanol fuel cells (DMFCs) appear as a promising next-generation fuel cells because the kinetics of methanol oxidation reaction (MOR) is faster than that for other alcohol fuels[12].

However, to reach sufficient reaction kinetics, an electro-catalyst must be used to improve the rate determining step of -COOH species formation in the MOR by reducing its activation barrier[13,14]. Palladium (Pd) metal is considered the prime candidate as MOR electrochemical catalyst, owing to its excellent stability, but the high cost and low activity towards MOR have restricted its potential application. To reduce the catalyst's cost and accelerate the sluggish kinetics, the synthesis of, multi-component Pd-based alloy catalysts has been investigated[15–18]. The integration of Sn[19], Ni[20], Cu[21], Ru[22], Rh[23], Ag[24], Er[25], or Au[8] modifies the structural and electronic properties of Pd, with potential to remarkably boost the MOR activity and durability[26].

Among these bi-metallic electrocatalysts, the solid-solution Pd-Au alloy catalyzes through a direct -OOH formation pathway[27] and exhibits superior activity in the selective oxidation of methanol[28–30]; therefore, it has been the focus of numerous studies[31,32]. Typically, a series of $Pd_xAu_y$ nano-catalysts with different Pd/Au atomic ratios were prepared by wet-chemical synthesis with $NaBH_4$ reductant[33–36] and their alcohol oxidation activity was subsequently compared.

For the development of AFCs, it is also necessary to understand the degradation mechanisms of the catalyst to facilitate accurate predictions of the electrocatalytic performance evolution and the associated fuel cell operational lifetime, as well as to guide the design of stable catalysts. For this, accelerated degradation studies must be undertaken.

Here, we synthesized a model system of $Pd_1Au_1$ solid-solution catalyst using $NaBH_4$ reduction[37]. All the chemicals purchased from Sigma Aldrich were high-purity level analytical grades. 0.05 M of the Pd-Au (molar ratio of 1:1) ions solution was prepared by simultaneously dissolving $HAuCl_4$ and $K_2PdCl_4$ precursors in deionized water. After sonication to form a homogenous solution, 0.5 M of $NaBH_4$ solution was added to the mixture to promote precipitation. A black powder was collected with a centrifuge, washed with distilled water for three times, and dried in a vacuum desiccator (see detailed synthesis in Supplementary Material).

Pd-Au nanoparticles were characterized by scanning electron microscopy (Zeiss Merlin), scanning transmission electron microscopy (STEM) (JEOL EM-2200FS) and energy-dispersive X-ray spectroscopy (EDS). Figure 1a and 1b show a complex 3D network of agglomerated nanoparticles forming ligaments with approximately 10 nm in width. This aerogel-like morphology arises from the rapid metal nanoparticle nucleation and agglomeration caused by the introduction of the strong reducing agent ($NaBH_4$) in the solution[38–40]. The crystalline structure determined by TEM and X-ray diffraction (XRD) (Rigaku Smartlab) is face-centered cubic. The XRD position of the (111) reflection appears at 39.06° while the (111) peak position of pure Pd and Au powders are respectively indexed at 39.44° and 37.34° confirming the formation of the Pd-Au solid solution phase. The STEM-EDS elemental mapping shows that Pd and Au elements are homogeneously distributed in the agglomerated particles, and no sign of dominant Pd or Au-rich solid-solution phase is observed.

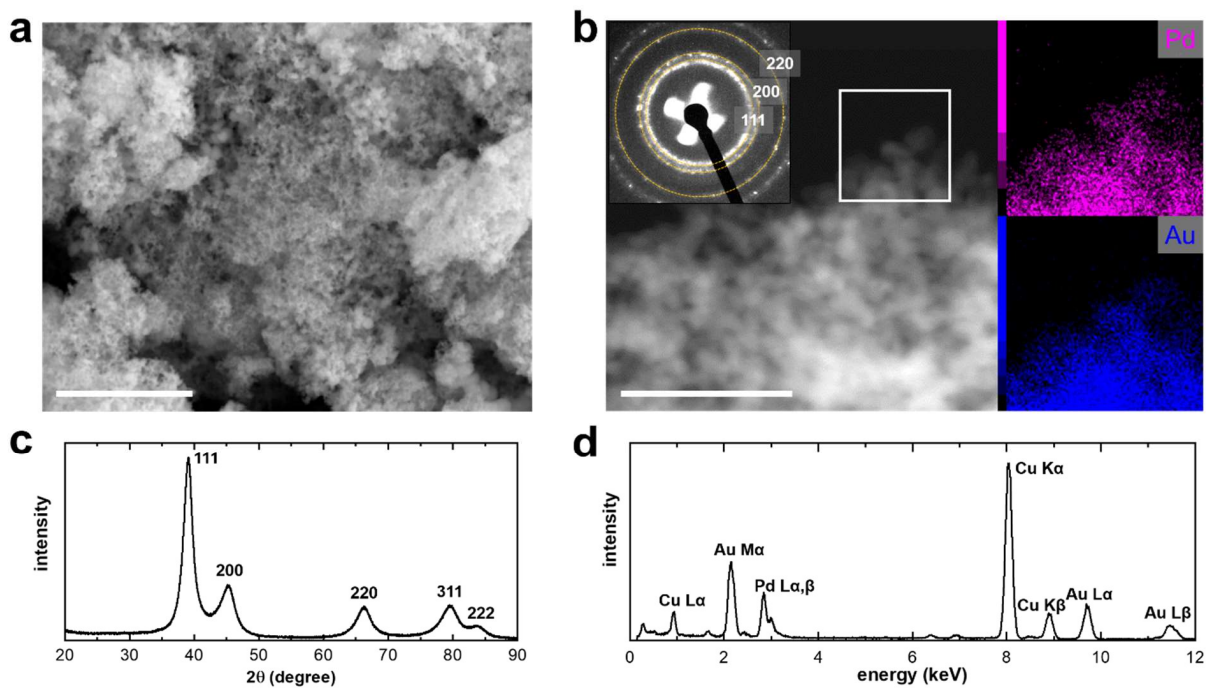

**Figure 1.** (a) SEM and (b) STEM images of the Pd-Au nanoparticles. Scale bars are 500 and 50 nm, respectively. Insets in (b) display the selected area diffraction pattern (SAED) and elemental mapping of Pd (purple) and Au (blue). (c) XRD pattern and (d) EDS spectrum of the Pd-Au nanoparticles. Cu (at 1 and 8 keV) and C (at 0.3 keV) are likely from a TEM grid and film, respectively.

We then performed atom probe tomography (APT) (CAMECA LEAP 5000 XS), and the 3D atom map for the PdAu nanoparticles embedded in a Ni matrix using the approach outlined in Ref.[41] (detailed sample preparation for APT is found in Supplementary Material) is shown in Figure 2a. A set of iso-composition surfaces delineating regions containing over 50 at. % of Pd+Au highlights the aerogel-like morphology. A substantial amount of B from the reducing agent, consistent with our previous studies on Pd-aerogels[42], is detected inside the material, as seen in Figure 2. Within the structure where most regions are PdAu, we can differentiate regions with two distinguished compositions, $Pd_3Au$ and PdAu, Figure 2c (i and ii), and they contain 0.235 ±0.077 and 2.406 ±0.244 B at.%, respectively. Each composition was measured from individual extracted region-of-interests with a volume of 10 × 10 × 10 nm$^3$. This ten-fold difference in B concentration can be attributed to smaller interstitial sites in the unit cell of Pd$_3$Au, considering that the lattice parameter for pure Pd is 0.389 nm and for pure Au is 0.408 nm. It is worth noting that Pd$_3$Au is a disordered phase because we did not observe any superlattice peaks in XRD and SAED patterns. Additionally, B is found as elemental boron (solid solution) in both regions, PdAu and Pd$_3$Au, due to the absence of oxygen (trace O only present in multi-junction grain boundaries), while B atoms are not located at O-rich regions, which could suggest borates (see Fig. S2).

Figure 2d displays the one-dimensional (1D) compositional profile normal to the particle/Ni matrix interface of the Pd$_3$Au and PdAu region showing no surface segregation. Both XRD and STEM-EDS results suggest the formation of a homogeneous Pd$_1$Au$_1$ solid-solution. In bulk Pd-Au, the mixing enthalpies for Pd and Au are negative and thus a pronounced phase separation is not expected[43]. However, the alloying behavior for high surface-to-volume-ratio particles is different[44]. In a Pd-Au surface phase diagram, which is calculated from surface energies as function of the corresponding compositions, the segregation energy of Pd on Au was found to be positive (0.15 eV/atom), whereas it was negative for Au on Pd (-0.14 eV/atom)[45]. This implies that locally Pd or Au atom has different mixing behavior when it is deposited on Au or Pd-rich surface. Additionally, from density functional theory calculation, the relative energy of Pd atom within an Au surface layer is higher than sub-surface or in the bulk, resulting in its preference to be alloyed inside the Au bulk[46]. These results suggest the possibility of phase separation of the Pd-Au solid solutions when the surface-to-volume ratio is high, *i.e.* as it would be in our nano materials. To the best of the authors' knowledge, the formation of two solid solution phases with different chemical compositions (Pd$_3$Au and PdAu) and the more intense B incorporation into the latter was never reported for such nano-materials[47].

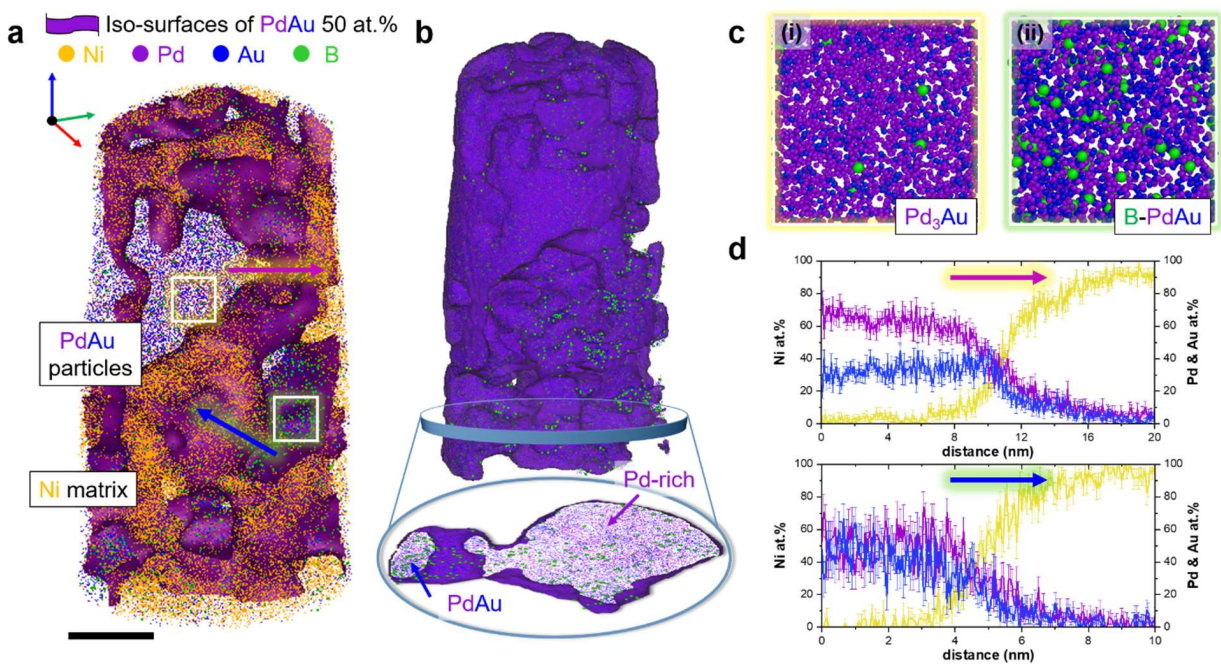

**Figure 2.** (a) Reconstructed 3D atom map of as-synthesized PdAu nanoparticles embedded within an electrodeposited Ni film. The scale bar is 10 nm. Yellow, purple, blue, and green mark the reconstructed position of Ni, Pd, Au, and B atoms, respectively. (b) Pd-Au iso-volume with a composition threshold of 50 at.% of Pd+Au. Bottom inset shows a 2-nm thin slice viewed along the measurement direction (z). (c) Extracted cuboidal atom maps (5 × 5 × 5 nm$^3$) of two different phases with different B amount. (d) 1D composition profile across one of the agglomerated Pd$_3$Au (up) and PdAu (down) particles of entire PdAu particles.

**Table 1.** Composition analysis of before and after MOR of as-synthesized Pd-Au nanoparticles from EDS and APT measurements.

| | Before MOR | | | | | | After MOR | | | |
|---|---|---|---|---|---|---|---|---|---|---|
| | EDS | | APT | | | | APT | | | |
| | SEM | STEM | surface | ROI (i) | ROI (ii) | Overall | surface | ROI (iii) | ROI (iV) | Overall |
| **Pd at.%** | 53.38 ±1.92 | 50.98 ±0.44 | 57.214 ±0.343 | 75.093 ±0.688 | 52.269 ±0.795 | 56.408 ±0.015 | 46.791 ±0.125 | 57.014 ±0.203 | 55.310 ±0.095 | 54.381 ±0.015 |
| **Au at.%** | 46.62 ±1.90 | 49.02 ±0.63 | 41.525 ±0.341 | 24.671 ±0.686 | 45.325 ±0.792 | 41.77 ±0.015 | 53.049 ±0.125 | 42.961 ±0.203 | 40.732 ±0.094 | 44.503 ±0.015 |
| **B at.%** | ND | ND | 1.262 ±0.077 | 0.235 ±0.077 | 2.406 ±0.244 | 1.823 ±0.004 | 0.160 ±0.010 | 0.024 ±0.006 | 3.958 ±0.037 | 1.116 ±0.003 |

In order to perform the electrochemical experiments, a three-electrode glass cell was employed, in which the working electrode was a drop-casted catalyst on a glassy carbon electrode and the counter electrode was a Pt wire. The electrode catalyst ink was prepared by mixing 30.8 mg of as-synthesized PdAu powder and 46.2 mg of carbon black (Sigma Aldrich) in a mixture of 2.0 mL of isopropyl alcohol, 2.0 mL of distilled water, and 10 μL of Nafion solution (Sigma Aldrich). Subsequently, the colloidal suspension was treated by ultrasonic wave agitation for 10 min for high dispersion. Then, the resultant ink (7 μL) was dropped onto a commercial glassy carbon electrode (Pine Instruments) and dried in an oven at temperature of 60 °C for 30 min. Methanol oxidation performance of as-synthesized PdAu catalyst was measured in a 1.0 M NaOH + 0.5 M methanol solution by cycling the potential between 0.25 and 1.2 V with scan rate of 100 mV/s.

In Figure 3, two anodic peaks are labelled, which are attributed to methanol oxidation at 0.85 V *vs*. RHE in the anodic scan (blue arrow) and adsorbed carbonaceous removal at 0.7 V *vs*. RHE in the cathodic direction (purple arrow)[48–50]. In order to investigate the chemical evolution of PdAu catalyst, accelerated degradation test (ADT) was then carried out, through sequential cycling of potential from 0.6 to 1.0 V *vs*. RHE in 1.0 M NaOH solution at a scan rate of 200 mV s$^{-1}$. After 1000 ADT cycles, the catalyst has partially lost activity, and the anodic peak current has dropped from 1.54 to 0.78 A cm$^{-2}_{geo}$, and the onset potential decreased from 0.497 to 0.522 V *vs*. RHE.

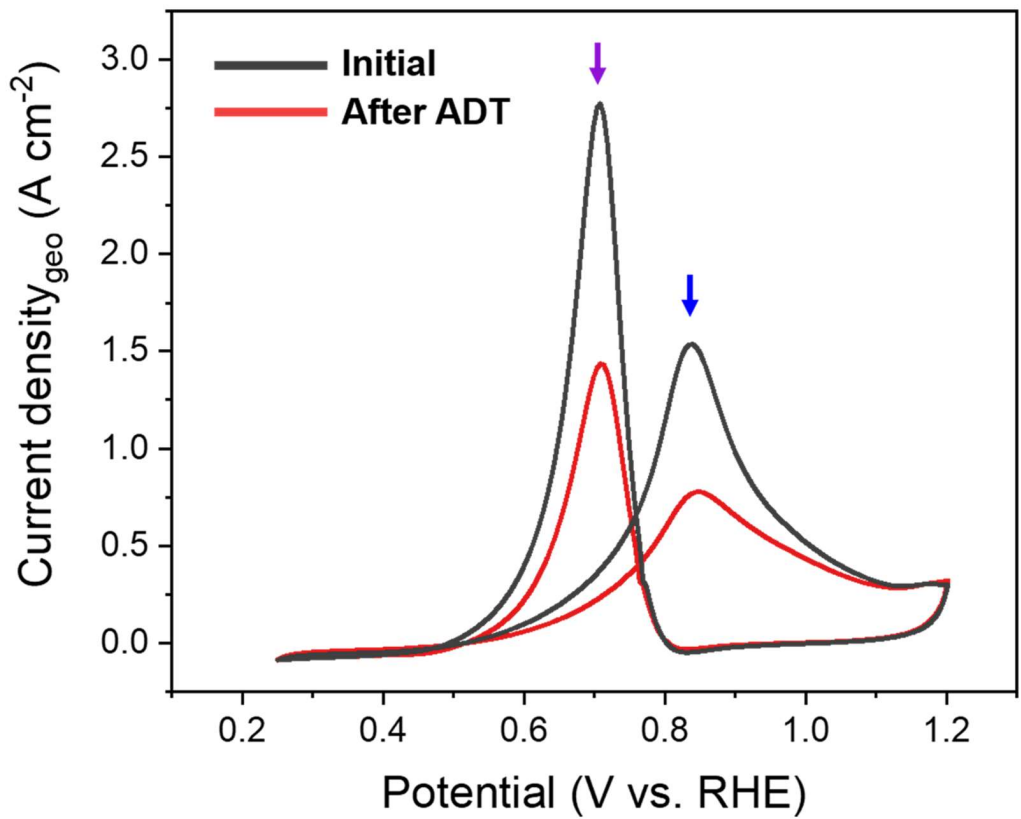

**Figure 3.** Cyclic voltammograms of methanol oxidation of Pd-Au sample before (black) and after (red) 1000 cycles of ADT. Blue and purple arrows mark the forward and reverse anodic peaks, respectively.

After ADT 1,000 cycles, we collected the catalysts from the working electrode and analyzed by APT. Figure 4a displays the 3D atom map of the degraded PdAu catalyst. The overall atomic composition has changed after ADT 1,000 cycles, such that the overall atomic ratio of Pd to Au has decreased from 1.35 to 1.22 (see Table 1). The Pourbaix diagram predicts that Pd metal likely dissolve as $Pd(OH)_4$ during anodic polarization in strong basic solutions (at 0.87 V *vs.* RHE[51] at pH 13), whereas for Au this reaction occurs at a relatively higher potential of 1.4 V *vs.* RHE at pH 13[52]. The potential range of the performed ADT was 0.6 to 1.0 V, leading to a preferential dissolution of Pd into the alkaline electrolyte. Both 1D compositional profile and proximity histogram were plotted normal to the aerogel-matrix interface[53], Figure 4b, and they reveal noticeable surface segregation of Au, which can be attributed to the potential-driven Pd dissolution. Generally, the activity of electrocatalysts is strongly linked to their surface configuration and composition, hence Au-segregation explain the drop in activity as Au electrodes have negligible activity toward MOR[54,55].

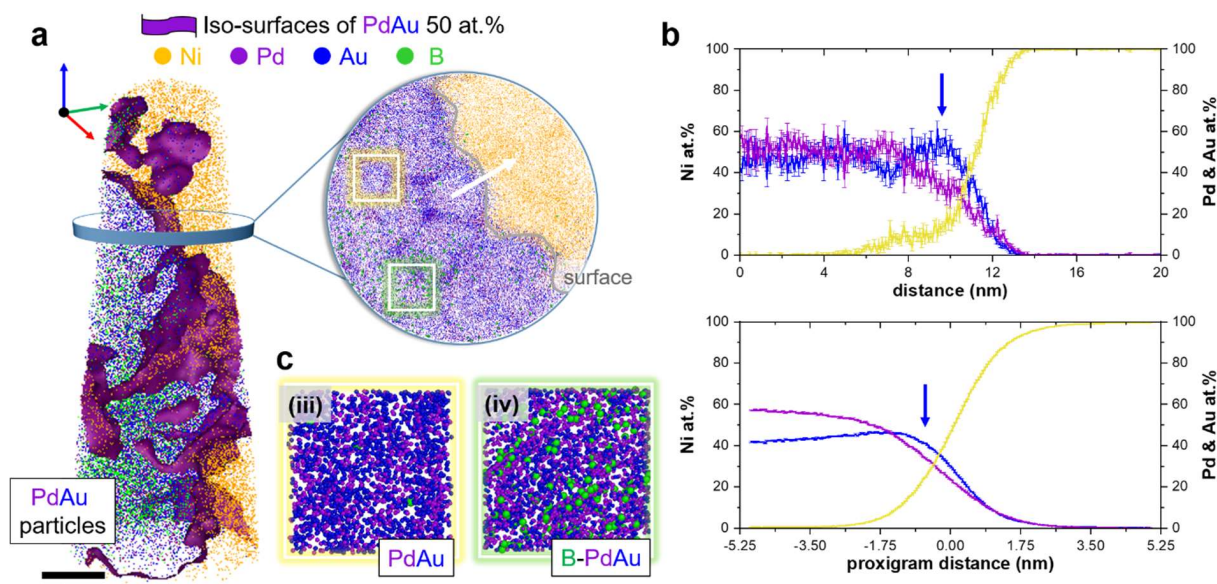

**Figure 4.** (a) Reconstructed 3D atom map of ADT-cycled PdAu nanoparticles embedded within an electrodeposited Ni film. A scale bar is 20 nm. (b) 1D composition profile across the interface (up) and proximity histogram of entire particles. (c) Extracted cuboidal atom maps ($5 \times 5 \times 5$ nm$^3$) of two different phases with different B amount.

Besides, the overall B content was reduced by a factor of ~2 after ADT 1,000 cycles. There are still regions of low and high B concentrations, Figure 4 (iii and iv), however, now they both have a 1:1 Pd to Au atomic ratio. In addition, no regions with a Pd$_3$Au stoichiometry were detected in the relatively large reconstructed APT dataset. This may indicate a yet unknown local electrochemically-driven transformation from Pd$_3$Au to Pd$_1$Au$_1$ owing to the leaching of Pd, besides the surface B leaching. In a previous report, B incorporation in a Pd nanoaerogel enhanced the activity towards the hydrogen evolution/oxidation reaction[42] by weakening the hydrogen binding energy of Pd. Similarly, the positive effect of B incorporation was also observed in PdIr catalysts activity towards methanol oxidation due to electronic/strain effects [56]. Thus, the leaching of B may hence also be modifying the binding energy of reactants resulting in an additional drop of activity after the ADT.

Despite the known degradation of PdAu catalysts upon operation of alcohol fuel cells[5,57], the local microstructural modifications responsible for such phenomenon are not well understood, which limits the possibilities for modelling of their evolution and operational lifetime predictions. Here, we used advanced characterization tool (e.g. APT) to reveal the following transformations responsible for the catalyst degradation: (1) Pd leaching from PdAu particles, triggering an electrochemical transformation from $Pd_3Au \rightarrow PdAu$ solid solutions, (2) the formation of Au-rich regions at the surface of the catalyst and (3) B leaching from both the surface of the particles and from the original $Pd_3Au$ regions. It is not clear which chemical modification is responsible for the degradation of the catalyst (and to which degree), but our study emphasizes the importance of understanding both the effect of B in Pd-Au catalyst and how to prevent surface Pd dissolution during MOR, in order to design superior catalysts.

**Acknowledgements**

L.S.A and S.-H.K. contributed equally on this work. S.-H.K. and B.G. acknowledge financial support from the German Research Foundation (DFG) through DIP Project No. 450800666. C.J. is grateful for financial support from the Alexander von Humboldt Foundation. S.Z. acknowledges funding from the DFG within the framework of SPP 2370 (Project number 502202153).

# Supplementary Material

# Revealing compositional evolution of PdAu electrocatalyst by atom probe tomography

Leonardo Shoji Aota[a], Se-Ho Kim[a,b,*], Chanwon Jung[a], Siyuan Zhang[a], Baptiste Gault[a,c,*]

[a] Max-Planck-Institut für Eisenforschung, Max-Planck-Straße 1, 40237 Düsseldorf, Germany
[b] Department of Materials Science and Engineering, Korea University, Seoul 02841, Republic of Korea
[c] Department of Materials, Imperial College London, SW7 2AZ London, UK
[*] Corresponding authors

# Methods

**PdAu aerogels synthesis**

PdAu metal aerogels were synthesized under ambient conditions using high-quality analytical grades from Sigma Aldrich. 0.163 g of $K_2PdCl_4$ (98 %) and 0.197 g of $HAuCl_4 \cdot 3H_2O$ (99 %) were added to 10 mL of deionized water. In a separate vial, 0.189 g of $NaBH_4$ (99 %) were dissolved in 10 mL of deionized water. After sonicating for 5 mins, the solutions were mixed to start the synthesis reaction. The mixture was kept still for 10 min. The final as-formed gels were collected with a centrifuge and washed with water for three times.

**PdAu co-eletrodeposition sample preparation for FIB (focused ion beam)**

The co-electrodeposition technique was used to prepare the PdAu nanoparticles for APT. A Ni ion electrolyte was produced by dissolving 15 g of nickel (II) sulfate hexahydrate (98%) and 2.25 g of citric acid (99.5%), both from Sigma-Aldrich, in 50 ml of distilled water. The PdAu nanoparticles were dispersed in the prepared Ni ions electrolyte via sonication for 20 min. Subsequently, the solution was poured on a vertical cell, consisting of a Cu substrate and a Pt counter electrode. A constant current of -38 mA was used for 500 s to deposit a 10-µm thick Ni film. The Ni deposition completely encapsulated the PdAu nanoparticles.

**Focused ion beam (FIB) sample preparation for APT**

APT samples were fabricated by FIB. Protrusions from the Ni+PdAu film indicate the presence of embedded nanoparticles (Fig. S1a). Once confirmed, the presence of void-free PdAu nanoparticles, a lift-out of the region of interest was produced. The lift-out was pasted on Si posts with the ion beam consolidation of Pt-C molecular compounds. Each post resulted in a single sample (Fig. S1b), which was then sharpened into a needle-shape APT specimen (Fig. S1c).

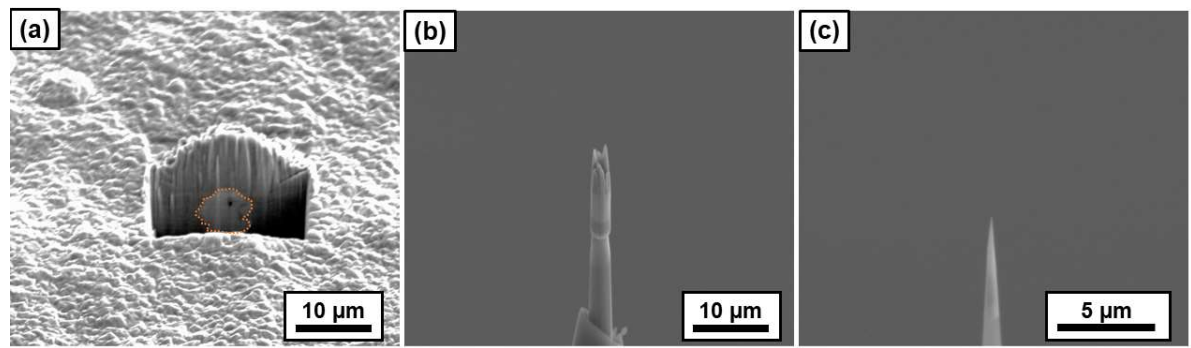

**Fig. S1** – (a) Cross section from the electrodeposited Ni film with a PdAu catalyst cluster (highlighted by orange dashed lines) below a surface protrusion and (b) the specimen pasted onto the Si post. (c) The sharpened specimen from PdAu particles/Ni for APT measurement.

**Atom probe tomography**

The final specimens of the PdAu/Ni sample were mounted in the atom probe analysis stage (CAMECA LEAP 5000XS). The measurement parameters were the stage temperature of 60 K, laser pulsing frequency of 125 kHz, detection rate of 0.5%, and laser energy of 60 pJ. The collected data were analyzed with commercial IVAS software developed by CAMECA.

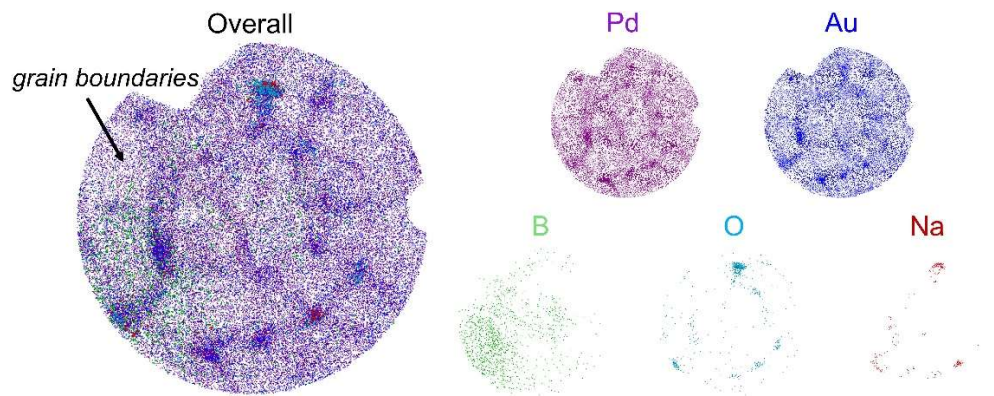

**Fig. S2.** Elemental distributions of Pd, Au, B, O, and Na from the 3-nm thin sliced ADT-cycled PdAu catalysts (from Fig. 4). Note that O and Na atoms are segregated at grain boundary junctions.